\documentclass{article}

\PassOptionsToPackage{square,numbers}{natbib}

 \usepackage[preprint]{neurips_2022}

\usepackage[utf8]{inputenc} 
\usepackage[T1]{fontenc}    
\usepackage{hyperref}       
\usepackage{url}            
\usepackage{booktabs}       
\usepackage{amsfonts}       
\usepackage{nicefrac}       
\usepackage{microtype}      
\usepackage{xcolor}         
\usepackage{graphicx}
\usepackage[skip=2pt]{caption}

\title{Compositional generalization through abstract representations in human and artificial neural networks}

\author{
    Takuya Ito\thanks{Work done as an intern at IBM Research AI} \\
    Yale University \\
    \texttt{taku.ito1@gmail.com}
 \And
 Tim Klinger \\
IBM Research AI \\
\texttt{tklinger@us.ibm.com} \\
\And
Douglas H. Schultz \\
University of Nebraska-Lincoln \\
\texttt{dhschultz@unl.edu} \\
\And
John D. Murray \\
Yale University \\
\texttt{john.murray@yale.edu} \\
\And
Michael W. Cole \\
Rutgers University \\
\texttt{michael.cole@rutgers.edu} \\
 \And
Mattia Rigotti \\
IBM Research AI \\
\texttt{mr2666@columbia.edu} \\
}

\begin{document}

\maketitle

\begin{abstract}
Humans have a remarkable ability to rapidly generalize to new tasks that is difficult to reproduce in artificial learning systems.
Compositionality has been proposed as a key mechanism supporting generalization in humans, but evidence of its neural implementation and impact on behavior is still scarce.
Here we study the computational properties associated with compositional generalization in both humans and artificial neural networks (ANNs) on a highly compositional task.
First, we identified behavioral signatures of compositional generalization in humans, along with their neural correlates using whole-cortex functional magnetic resonance imaging (fMRI) data.
Next, we designed pretraining paradigms aided by a procedure we term {\em primitives pretraining} to endow compositional task elements into ANNs.
We found that ANNs with this prior knowledge had greater correspondence with human behavior and neural compositional signatures.
Importantly, primitives pretraining induced abstract internal representations, excellent zero-shot generalization, and sample-efficient learning.
Moreover, it gave rise to a hierarchy of abstract representations that matched human fMRI data, where sensory rule abstractions emerged in early sensory areas, and motor rule abstractions emerged in later motor areas.
Our findings give empirical support to the role of compositional generalization in human behavior, implicate abstract representations as its neural implementation, and illustrate that these representations can be embedded into ANNs by designing simple and efficient pretraining procedures.
\end{abstract}

\section{Introduction}

Humans can efficiently transfer prior knowledge to novel contexts, an ability commonly referred to as transfer learning.
One proposed mechanism underlying transfer learning is compositional generalization (or compositional transfer) -- the ability to systematically recompose learned concepts into novel concepts (e.g., ``red'' and ``apple'' can be combined to form the concept of a ``red apple'') \citep{cole_rapid_2012,frankland_concepts_2020,hupkes_compositionality_2020}.
Indeed, it has been suggested that an algorithmic implementation of compositional generalization is one of the key missing ingredients that ANN models need in order to achieve human-like learning and reasoning capabilities \citep{lake_building_2017,lake_generalization_2018}.
Therefore, quantifying how compositional generalization is manifested in human behavior and investigating its underlying implementation in biological brains is a natural first step to harness and deploy it in machine learning models. 

Recent studies that investigated compositionality in machine learning have typically relied on architectures comprised of specialized modules.
For instance, disentangled representation learning separates the independent factors underlying the structure of the input data into disjoint components of the feature vector \citep{higgins_towards_2018,higgins_unsupervised_2021,montero_role_2020,higgins_beta-VAE_2016}.
Program synthesis methods achieve state-of-the-art performance on systematic generalization \citep{hupkes_compositionality_2020} through model architectures built by combining specialized neural and symbolic program modules interacting to search over a space of valid production rules \citep{lake_compositional_2019, nye_learning_2020, ruis_benchmark_2020}.

Complementing these studies, \emph{abstract representations} have been recently proposed as vector representations that reconcile compositional generalization with distributed neural codes \citep{bernardi_geometry_2020}. 
In particular, \emph{parallel abstract representations} -- representations with a high Parallelism Score as previously defined \citep{bernardi_geometry_2020} -- support out-of-context generalization by encoding changes in individual variables as a linear shift in the representations. 
This notion of abstraction implies that these representations are compositionally additive; novel compositions are encoded as the vector sum of distinct abstract representations.
This is similar to how word2vec embeddings solve relational analogy tasks \citep{mikolov2013, levy2014} and generalizes disentangled representations by allowing for arbitrary affine transformations of disentangled codes.
Crucially, this type of representation is operationally defined in a way that can be quantified in neuroimaging data by computing the Parallelism Score metric defined in \citep{bernardi_geometry_2020}. 
In other words, parallel abstract representations are a computationally promising candidate as neural substrate implementing compositional generalization, and are also measurable in the human brain by computing the Parallelism Score across fMRI voxels during neuroimaging experiments.

This work is motivated by the working hypothesis that parallel abstract representations support compositional generalization.
Accordingly, we first characterized the behavioral signatures of compositional generalization in a task that systematically varied rule conditions across 64 contexts, showing that humans generalize better to tasks with greater similarity structure to previous tasks.
We then analyze fMRI imaging data showing that parallel abstract representations are distributed across the entire cortex in a content-specific way during the execution of the compositional task. 
This supports our working hypothesis that parallel abstract representations may implement compositional generalization in humans.
To test this hypothesis in ANNs, we designed a pretraining paradigm for ANNs to emulate humans' prior knowledge about the compositional task elements, finding that ANNs pretrained in this way exhibit 1) more abstract representations, 2) excellent generalization performance, and 3) sample-efficient learning.
This finding demonstrated that the degree of abstraction (induced through pretraining) directly impacted zero-shot compositional generalization performance in ANNs.
Finally, we find that the layerwise organization of abstract representations in pretrained ANNs recapitulates the content-specific distribution across the human sensory-to-motor cortical hierarchy.
Together, these findings provide empirical evidence for the role of abstract representations in supporting compositional generalization in humans and ANNs.

\subsection{Related work}
Several recent studies in neuroscience have applied analytic tools to identify the neural basis of rapid generalization in biological neural networks.
Such studies employed various measures -- cross-condition generalization \citep{bernardi_geometry_2020,reverberi_compositionality_2012,flesch_orthogonal_2022,cole_rapid_2011}, state-space projections of task-related compositional codes \citep{yang_task_2019,riveland_neural_2022,johnston_abstract_2021}, and Parallelism Score  \citep{bernardi_geometry_2020} -- to quantify the generalizability and abstraction of representations. 
Prior work in neuroscience has primarily evaluated compositionality in limited context settings (e.g., up to 10 contexts), or without manipulating different types of features (e.g., higher-order vs. sensory/motor features). 
Moreover, these neuroscience studies used simple task paradigms due to limitations in either the model organism (rodents and monkeys are unable to perform complex tasks \citep{bernardi_geometry_2020}) or to isolate specific types of abstraction in humans (e.g., logical abstractions \citep{reverberi_compositionality_2012}).
Here we significantly expand on prior work by using a 64-context compositional task that systematically varies different types of task features (e.g., sensory, motor, and logical rules) to evaluate content-specific abstractions across the entire brain and multilayer ANNs.
This work also complements related work in compositional generalization in machine learning \citep{lake_generalization_2018,ruis_benchmark_2020,hupkes_compositionality_2020,lake_compositional_2019,rosenbaum_routing_2019,yang_dataset_2018,hudson_gqa_2019}.
However, those studies primarily focused on building models that improve on current compositional generalization benchmarks on arbitrarily complex compositional tasks, such as SCAN \citep{lake_generalization_2018}, COG \citep{yang_dataset_2018}, or GQA \citep{hudson_gqa_2019}.
Importantly, these studies did not directly benchmark ANN behavior (or representations) against human behavioral and neural data, making a direct comparison difficult.
Here we leveraged a non-trivial 64-context compositional paradigm to investigate the representations that facilitate compositional generalization in both humans and ANNs.


\section{Methods}

\subsection{The Concrete Permuted Rule Operations (C-PRO) task paradigm}

\begin{figure}[h]
  \includegraphics[width=\linewidth]{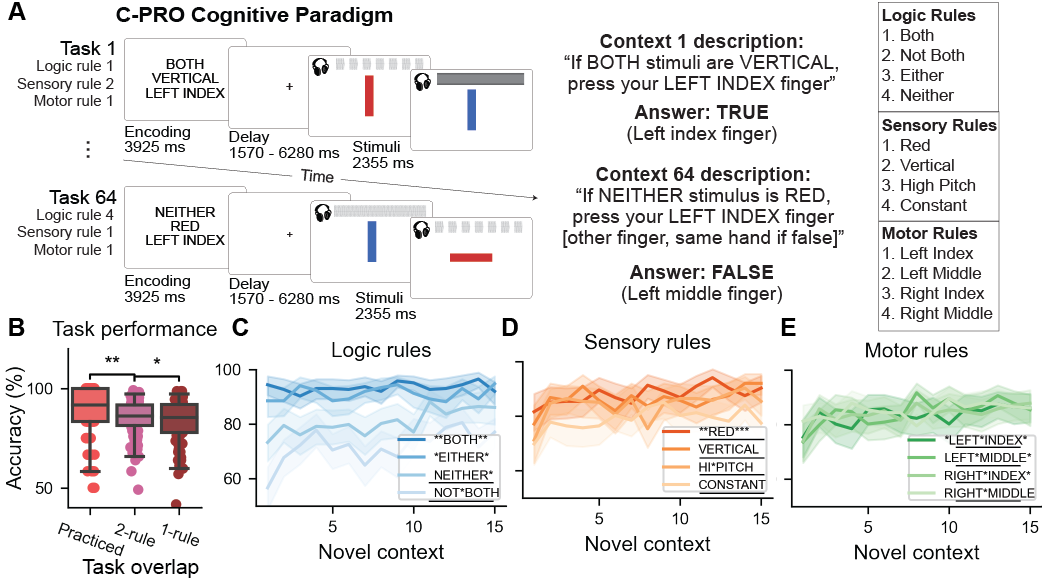}
  \caption{a) The C-PRO paradigm permutes 12 rules belonging to three different rule domains -- logical, sensory, and motor gating -- to generate up to 64 unique contexts. b) Human performance on novel task contexts was significantly lower than on practiced contexts (participants were trained on four practice contexts prior to the test session). Moreover, subjects performed novel task contexts with more rule overlap with practiced contexts at a higher accuracy. c-e) Task performance as a function of task trials for each rule (novel contexts only). Consistent with compositional generalization, participants had a significant increase in task performance in 10/12 rules, even though each rule was used in a novel context. Shaded area around line plots (c-e) reflects the 95\% confidence interval.}
   \label{fig:fig1}
\end{figure}

We used the C-PRO paradigm (Fig. \ref{fig:fig1}a) during fMRI acquisition and ANN model training. 
Briefly, the C-PRO paradigm permutes specific task rules from three rule domains (logical decision, sensory semantic, and motor response) to generate dozens of novel task contexts.
This creates a context-rich dataset in the task context domain.
The sensory rule indicates which stimulus feature should be attended to.
The logic rule specifies a Boolean operation to be implemented on the stimulus feature set. 
The motor rule specifies a specific motor action (i.e., a button press with a specific finger).
One of 256 possible unique stimulus combinations could be presented with each task context.
Visual dimensions included either horizontal or vertical bars with either blue or red coloring.
Auditory dimensions included continuous (constant) or beeping high or low pitched tones.

Each rule domain (logic, sensory, and motor) consisted of 4 specific rules (Fig. \ref{fig:fig1}a).
A task context is comprised of one rule from each domain, for a total of 64 possible task contexts (4 logic x 4 sensory x 4 motor).
Subjects were trained on 4/64 ``practiced'' contexts prior to the fMRI session.
The 4 practiced contexts were selected such that all 12 rules were equally practiced.
Subjects' mean performance across all trials was 84\% (median=86\%; chance=25\%). See Appendix A.2 for details\footnote[1]{The fMRI dataset is publicly available here: \url{https://openneuro.org/datasets/ds003701}}. 

\subsection{The geometry of abstract neural representations}

Behavioral signatures of compositional generalization can be investigated by measuring behavioral performance as a function of task composition and prior learning.
Neural signatures of generalization can be identified using analysis methods that characterize the geometry of neural activations during task generalization.
In particular, prior work proposed the Parallelism Score (PS) \citep{bernardi_geometry_2020} as a measure to evaluate the consistency of task variable representation across contexts. 
Intuitively, PS identifies a coding axis (i.e., the parallel displacement vector) across task contexts that aids generalization.

\begin{figure}[h]
  \includegraphics[width=\linewidth]{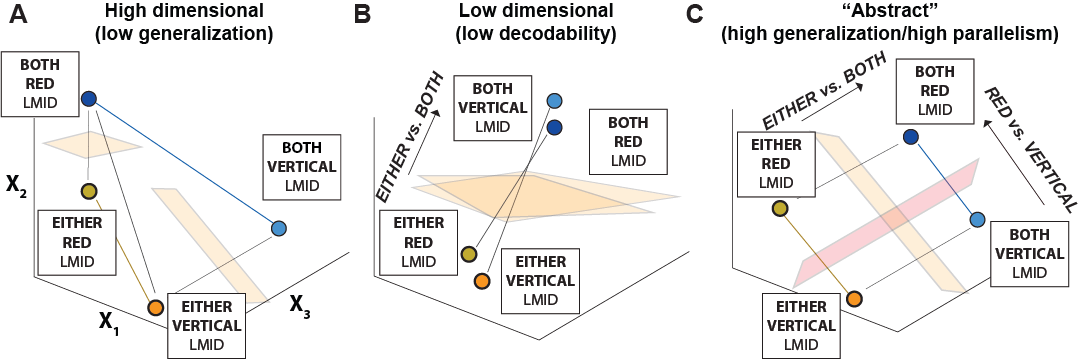}
  \caption{Hypothetical geometric configurations of neural activation space for ``BOTH vs. EITHER'' and ``RED vs. VERTICAL'' rule contrasts. a) High-dimensional representations of task activations lead to low PS (in addition to low generalizability across conditions) of rules. b) Low-dimensional representations lead to overall low decodability, but some generalizability (across limited features). c) Parallel Abstract Representation of the neural activations leads to high generalizability.}
   \label{fig:fig2}
\end{figure}

We posit that representations with high PS (the specific type of abstract representation we investigate) support compositional generalization in human behavior.
We illustrate here how PS is reflected in the geometry of neural representations with respect to the rule domains of the C-PRO task.
Let us consider a set of C-PRO contexts with logic rules BOTH or EITHER, and sensory rules with values RED or VERTICAL (Fig.\ \ref{fig:fig2}). 
High PS in the logic rule domain indicates that the difference in activation vectors between contexts with BOTH and EITHER rules is the same when paired with either the RED or VERTICAL sensory rules.
Thus, a change from BOTH to EITHER results in the same parallel change irrespective of the sensory rule (Fig.\ \ref{fig:fig2}c).
In contrast to unstructured high-dimensional representations (Fig.\ \ref{fig:fig2}a, and see \citep{rigotti_importance_2013,fusi_why_2016}), this would afford high generalization, since the effect of changing the logic rule in either sensory rules automatically transfers to the other sensory rule.

\subsection{Parallelism score}

\begin{figure}[h!]
  \includegraphics[width=\linewidth]{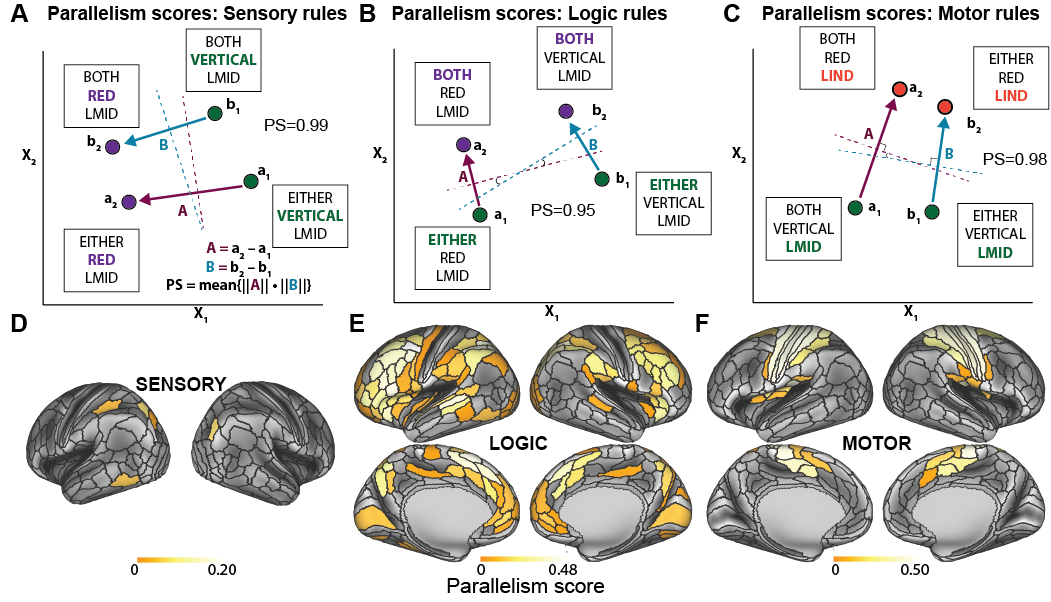}
  \caption{
  a-c) 2-D schematic visualization of PS estimation for the a) sensory, b) logic, and c) motor rule domains for a specific rule pair (e.g., RED vs. VERTICAL).
  Intuitively, PS captures the geometry of the neural activation space (ANN or fMRI data) by measuring the cosine angle between two linear decoders trained to distinguish two rule conditions in different task contexts. 
  d-f) PS was calculated for each rule domain for every brain region \citep{glasser_multi-modal_2016}. PS was highest in association areas for logic rules, dorsal attention network regions for sensory rules, and somatomotor network for motor rules.}
   \label{fig:fig3}
\end{figure}

We generalize the definition of PS by \citep{bernardi_geometry_2020} to tasks where variables can assume an arbitrary number of values (as opposed to being binary) and applied it to human fMRI and internal ANN activations.
PS is defined as the cosine angle of the coding directions of the same rules in different contexts in the neural activation space (e.g., voxels or neurons within a brain region) (Fig. \ref{fig:fig3}a-c).
A cosine angle close to 1 indicates coding directions that are highly parallel, despite differences in context.
We compute the coding angle for a specific rule dichotomy (e.g., the coding direction ``BOTH'' vs. ``EITHER'') by identifying all pairs of task contexts that had exactly the same secondary (sensory) and tertiary (motor) rules.
For each pair, we subtracted the fMRI voxel activation vectors associated with each context to obtain the vector that represented that coding direction (see Fig. \ref{fig:fig3}a). 
We did this for all other pairs in that coding direction. 
Defining $v_i$ as this coding vector for the $i$th pair, we computed the PS score for one dichotomy as $PS_{k} = \frac{1}{16} \sum_{i \neq j}^{16} cos(v_{i},v_{j}))$, since there are 16 possible pairs for each rule dichotomy within the C-PRO task. 
To obtain the PS for a specific rule domain (e.g., logic, sensory, or motor rules), $PS_{k}$ is computed for every coding direction, then averaged (e.g., for logic PS, the average of ``BOTH'' versus ``EITHER'', ``BOTH'' versus ``NEITHER'', etc.).

Statistical testing was performed using a non-parametric procedure, where we shuffled labels within each rule domain 1000 times and re-calculated PS to produce a null distribution. We corrected for multiple comparisons (across brain regions) using non-parametric family-wise error correction \citep{nichols_nonparametric_2001}.

\subsection{ANN construction and training}

The primary ANN architecture had two hidden layers (128 units each) and an output layer that was comprised of four units that corresponded to each motor response (Fig. \ref{fig:figs_annarch}; see Appendix section A.7 for additional details). 
Training used a cross-entropy loss function and the Adam optimizer \citep{kingma_adam_2017}. 
The ANN transformed a 28-element input vector into a 4-element response vector with the equation $ Y = f_{ReLU}(X_{h}W_{h}+b_{h}) $.
Weights and biases were initialized from a uniform distribution $U(-\sqrt{1/k},\sqrt{1/k})$, where $k$ is the number of input features from the previous layer.

\begin{figure}[h!]
  \includegraphics[width=\linewidth]{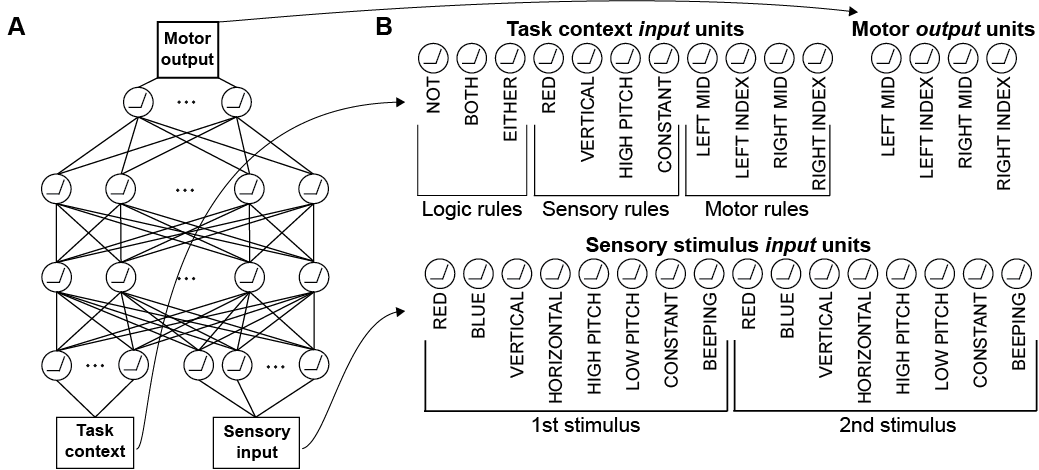}
  \caption{a) ANN architecture. We employed a simple architecture with two hidden layers (128 units each) with rectified linear units. b) The ANN's input and output units. Task input units were modeled as one-hot vectors for each rule across all rule domains. Sensory stimulus input units were modeled as categorical one-hot vectors for each stimulus type. Separate vectors were modeled for first and second stimulus presentations. Motor outputs had a unit corresponding to each motor response. Noise was injected with input features when calculating PS to reduce the orthogonality of input features.}
   \label{fig:figs_annarch}
\end{figure}

Training on the C-PRO task was performed in a sequential learning paradigm. To mimic the human experiment, an arbitrary set of four practiced contexts was initially selected for training. (This was randomly selected across different ANN initializations.) Then, novel task contexts were incrementally added into the set of training contexts.

\section{Results}

\subsection{Behavioral signatures of rapid compositional generalization in humans}

We evaluated human behavioral compositional generalization by assessing performance on novel contexts in the C-PRO paradigm. Since adult humans have decades of prior knowledge, subjects were able to compositionally generalize to novel task contexts without any training (novel accuracy=84.17\%, chance=25\%, Wilcoxon signed-rank p$<$0.0001). However, subjects performed the four practiced contexts better than novel contexts (practiced=87.67\%, novel=84.17\%;  p=0.003). We next assessed how performance on novel contexts changed as a function of shared rule structure to the practiced contexts. Consistent with compositional transfer of previously learned rules, performance on novel task contexts improved as a function of similarity to the practiced contexts (accuracy, 2-rule overlap=84.86\%; 1-rule overlap=83.48\%; practiced vs. 2-rule overlap, p=0.008; 2-rule vs. 1-rule overlap, p=0.03; Fig. \ref{fig:fig1}b). Though our findings are consistent with compositional transfer, we found that rapid transfer to novel contexts is more difficult. However, we found that increased exposure to specific rules improved performance on subsequent novel contexts using that same rule (all except for the ``Both'' and ``Either'' rules, likely due to ceiling effects, FDR-corrected p$<$0.05; Fig. \ref{fig:fig1}c-e). This suggests that even though performance in novel contexts is worse than practiced contexts, subjects can improve rule transfer with increased practice (or pretraining).

\subsection{Spatial and content-specific topography of abstract representations in human cortex}

We extended prior work to identify abstract representations using PS across the entire human cortex \citep{bernardi_geometry_2020}. 
We calculated PS for each rule domain separately (Fig. \ref{fig:fig3}a-c) using the vertices/voxels within each parcel (i.e., brain region) as activation vectors. We found topographic differences of sensory, logic, and motor rule abstractions tiled across human cortex (Fig. \ref{fig:fig3}d-f). Specifically, we found that statistically significant sensory rule abstractions were primarily identified in higher order visual areas and the dorsal attention network (i.e., brain areas involved in the top-down selection of visual stimuli) (PS of significant regions=0.15; family-wise error (FWE)-corrected p$<$0.05; Fig. \ref{fig:fig3}d). Logic rule abstractions were more widely distributed, but primarily observed in frontoparietal areas (PS of significant regions=0.22; FWE-corrected p$<$0.05; Fig. \ref{fig:fig3}e). Motor rule abstractions were primarily localized to somatomotor cortex (PS of significant regions=0.29; FWE-corrected p$<$0.05; Fig. \ref{fig:fig3}f). Notably, regions with abstract representations form a subset of regions of those that contain rule information using standard decoding methods (SFig. \ref{fig:figs_standarddecoding}). This ensures that high PS is accompanied by highly decodable representations (i.e., high dimensionality).

\subsection{Embedding prior knowledge into ANNs with simple pretraining tasks}

Human behavioral data suggested improved compositional generalization with increased task rule exposure, in addition to the years of ``pretraining'' from ordinary development (i.e., at least 18+ years). Thus, we sought to evaluate whether embedding prior knowledge of rules could improve compositional generalization in ANNs, while simultaneously investigating how prior knowledge impacts the geometry of ANNs' internal task representations. Given that the C-PRO task was specifically designed as a compositional task that conjoined three task rules, we created pretraining paradigms designed to teach ANNs basic rule knowledge (Fig. \ref{fig:fig4}; see Appendix for full description).

\begin{figure}[h]
  \includegraphics[width=\linewidth]{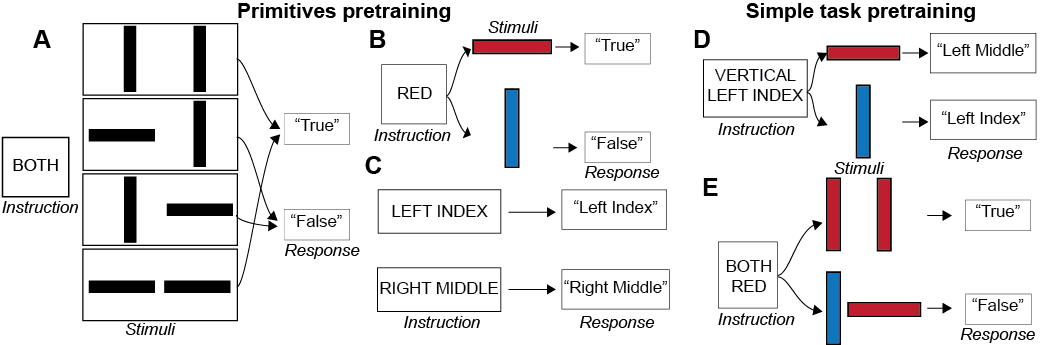}
  \caption{a) The logic rule primitives task involved teaching boolean relations among different logical operations. For example, when presented with the ``BOTH'' rule, the task was to distinguish two identical (``True'') versus two different (``False'') stimuli (i.e., same vs. different). b) Sensory rules involved mapping sensory rules onto stimulus features. c) Motor rules involved mapping motor rules onto motor output units. d-e) Simple task pretraining (2-rule tasks) was designed to teach the model how to perform simple (d) sensorimotor mappings and (e) logical-sensory gatings.}
   \label{fig:fig4}
\end{figure}

We constructed a simple feedforward ANN with two hidden layers (Appendix A.7; Fig. \ref{fig:figs_annarch}). 
This made it easier to investigate the effects of pretraining on internal representations, rather than architectural choices. 
We designed two pretraining paradigms: Primitives (1-rule) and Simple task (2-rule) pretraining. Primitives pretraining trained on 1-rule tasks that focused explicitly on learning the semantics of primitive rule features (Fig. \ref{fig:fig4}a-c). 
This included distinguishing sensory stimuli, learning motor response mappings (e.g., ``left index'' rule would lead to a left index response), and abstract logical relations, which involved learning the boolean relations amongst logic rules. 
Simple task pretraining focused on learning 2-rule conjunctions (i.e., a sensory and motor rule pairing / logical and sensory rule pairing) (Fig. \ref{fig:fig4}d-e). 
Importantly, these pretraining paradigms focused on learning primitive 1- or 2-rule associations that were significantly simpler than the full C-PRO task (3-rule combination).

\subsection{Pretraining induces abstractions, zero-shot performance, and sample efficiency}
\begin{figure}[h]
  \includegraphics[width=\linewidth]{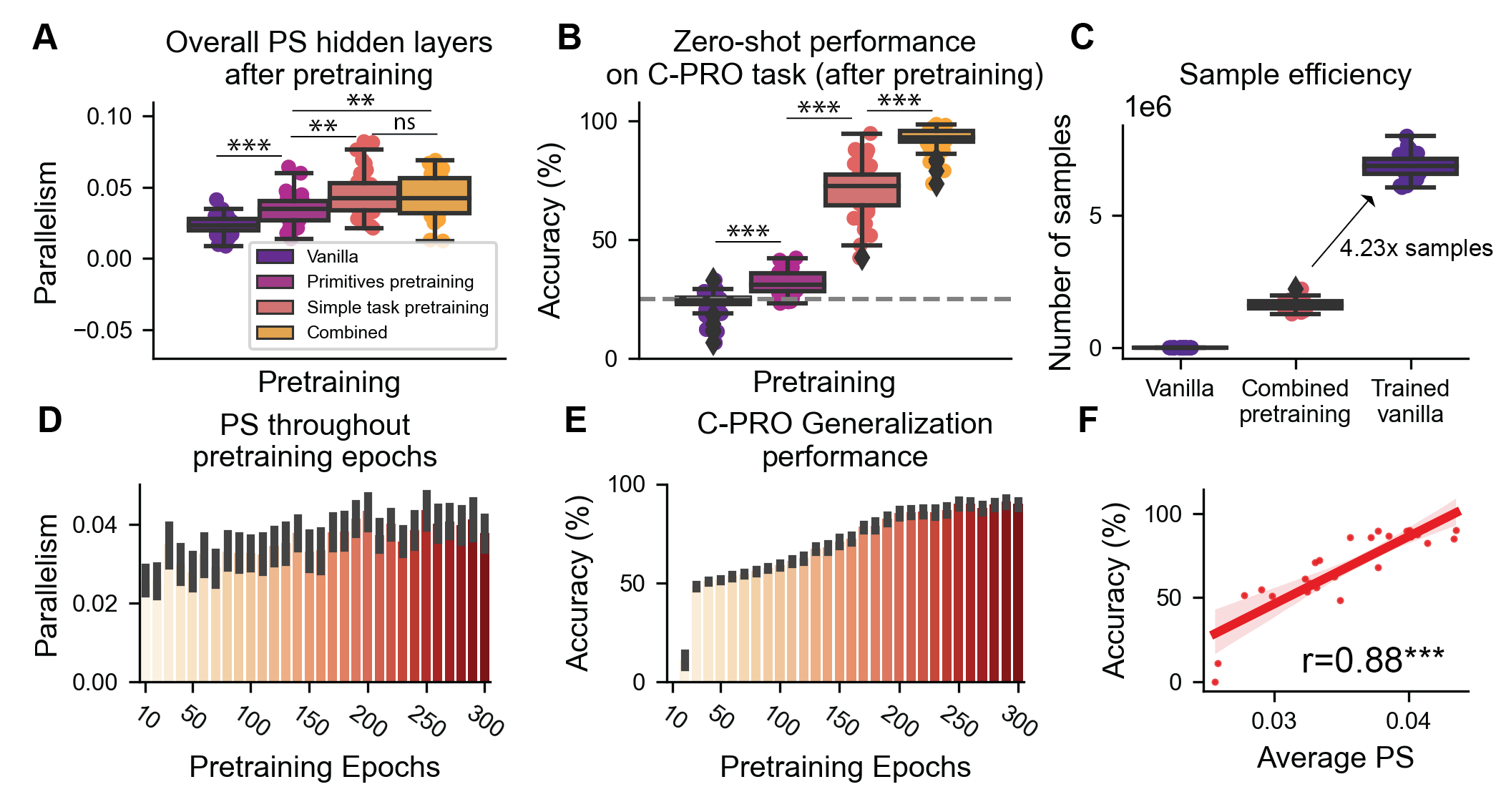}
  \caption{a) PS of hidden units averaged across all rule domains using noisy inputs. 
  b) Zero-shot learning of all 64 C-PRO contexts. 
  c) Sample efficiency of models (Combined and trained vanilla model were performance-matched). Total samples, including pretraining samples (if applicable).
  d) We computed the average PS of the ANN hidden layers throughout Combined pretraining. (Note that PS values are low due to input features containing large amounts of noise.)
  e) Zero-shot performance to the unseen C-PRO trials throughout pretraining. 
  c) The generalization performance and the PS scores were highly correlated with each other throughout pretraining, indicating a close link between abstract representations and improved task generalization.}
   \label{fig:fig5}
\end{figure}

We measured the PS in ANNs trained with different pretraining routines: Vanilla (no pretraining), Primitives pretrained, Simple task pretrained, and Combined (Primitives + Simple task pretrained). 
Each pretraining condition was trained until ANNs achieved 99\% accuracy (see A.8 for details).
PS was calculated for each rule domain using the ANN's hidden layer activations (see A.9). Pretrained ANNs had higher PS than the Vanilla ANN (Primitives vs. Vanilla, t(37)=5.26, p=1e-05; Simple task vs. Vanilla, t(37)=8.46, p=1e-11; Combined vs. Vanilla, t(37)=3.03, p=0.003) (Fig. \ref{fig:fig5}a). 
Moreover, PS increased from Primitives to Simple task pretraining (t(37)=3.91, p=0.0002), though no significant increase in PS was observed in Combined vs. Simple task pretraining.

We next evaluated the zero-shot performance on the full C-PRO task after pretraining (Fig. \ref{fig:fig5}b). 
As expected, the Vanilla ANN performed near chance (acc=23.25\%, chance=25\%, one-sided t(38)=-2.17, p=0.98). 
Primitives pretraining marginally improved zero-shot performance (acc=31.51\%, t(38)=8.09, p<1e-9). 
While Simple task pretraining exhibited significant improvement over Primitives pretrained models (acc=70.57\%, Simple task vs. Primitives, t(37)=19.84, p$<$1e-31), Combined pretraining had excellent zero-shot performance on the entire C-PRO task (acc=92.15\%, Combined vs. Simple task pretraining, t(37)=10.85, p$<$1e-16). 

We next sought to assess the impact of pretraining on learning/sample efficiency. 
We therefore trained a Vanilla network (no pretraining) on 60/64 C-PRO contexts to match the zero-shot performance of the Combined pretraining model (i.e., at least 90\% accuracy on the 60 context training set). We found that on the remaining test set (4/64 C-PRO contexts), the Vanilla trained model achieved 96.02\% generalization performance, but required up to 4.23x training samples to match the performance of the Combined model (Fig. \ref{fig:fig5}c). Critically, the 4.23x more training samples included all possible samples (pretraining and C-PRO samples). 
Thus, pretraining afforded both zero-shot generalization and sample efficient learning.

Finally, we measured PS and generalization performance throughout Combined pretraining. 
We found that PS and zero-shot performance increased with pretraining (Fig. \ref{fig:fig5}d,e), and were highly correlated (r=0.88; p$<$10e-9; Fig. \ref{fig:fig5}f). This illustrated that the abstract representations learned during pretraining directly facilitated zero-shot generalization, and is consistent with prior work demonstrating that the dimensionality of hidden representations is altered throughout training \citep{recanatesi_predictive_2021}.

\subsection{ANN pretraining leads to human-like compositional generalization}

\begin{figure}[h]
  \includegraphics[width=\linewidth]{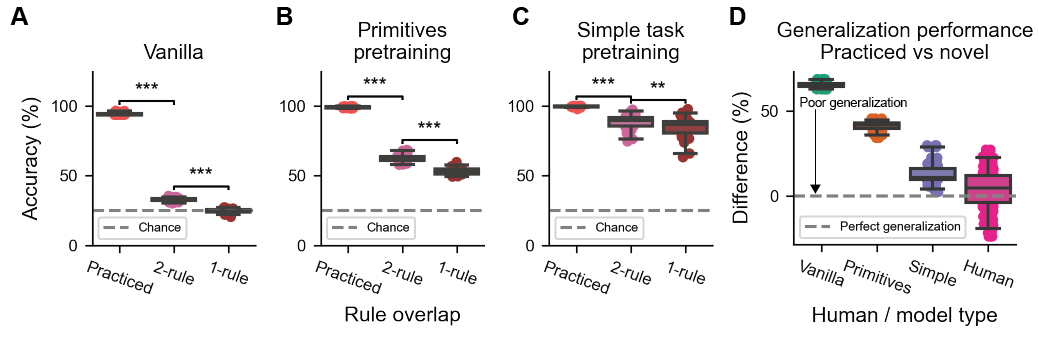}
  \caption{a-c) We trained the ANN architecture on 4/64 C-PRO task contexts with (a) a Vanilla ANN with randomly initialized weights, (b) an ANN after Primitives pretraining, and (c) an ANN after Simple task pretraining. Training on the 4 practiced contexts was stopped after the model achieved 90\% accuracy on those contexts. As in our empirical human data (Fig. \ref{fig:fig1}b), the ANN performed best on contexts in which it had previously seen (Practiced), followed by contexts with two overlapping (2-rule) and one overlapping (1-rule) rules. d) Generalization performance of ANNs with Primitives and Simple task pretraining was most similar to human generalization performance in novel contexts. (Lower values indicate better generalization.)}
   \label{fig:figs_anncompositionality}
\end{figure}

We evaluated the learning and generalization dynamics of ANNs with and without pretraining, after training ANNs on 4 of the full C-PRO contexts.
This matched the human experiment, since humans were exposed to 4 ``practice'' contexts prior to performing the remaining 60 novel contexts (Fig. \ref{fig:fig1}b).
(ANN training was stopped after achieving 90\% performance on the 4 practiced contexts.) 
We found overall poor generalization on novel task contexts in the Vanilla model (accuracy, practiced=94.37\%, novel=28.79\%; p$<$0.0001; Fig. \ref{fig:figs_anncompositionality}a). 
This suggested that ANNs with no prior knowledge cannot compositionally generalize. 
We subsequently compared generalization performance on ANNs after pretraining. 
We found that with Primitives pretraining, generalization performance significantly improved (57.97\%; Fig. \ref{fig:figs_anncompositionality}b). 
We observed additional improvements with Simple task pretraining (86.79\%; Fig. \ref{fig:figs_anncompositionality}c), achieving human-like generalization performance (Fig. \ref{fig:figs_anncompositionality}d). 

We next incrementally trained all ANN models on novel contexts, by adding one novel context into the training set at a time. 
We tested generalization performance on the held-out (test set) contexts until ANNs were trained on 63/64 contexts (SFig. \ref{fig:fig6}a). 
Generalization performance on novel contexts was significantly higher in ANNs with either pretraining routine (SFig. \ref{fig:fig6}b). 
This was despite the fact that all ANNs had the same stopping criteria (i.e., 90\% accuracy on the C-PRO training set). 
We ran an additional experiment where each of the ANNs were shown a fixed number of C-PRO task samples during training, replicating our core finding (SFig. \ref{fig:figs_ann_fixedepoch}). 
These findings suggest that the inductive biases formed during pretraining significantly improve downstream generalization performance.

\subsection{Pretraining ANNs facilitates sample-efficient learning throughout novel task learning}

We sought to evaluate how pretraining impacted sample efficiency. We found that pretrained ANNs became more sample efficient as the training set expanded, even after accounting for total number of (pretraining and C-PRO) samples (SFig. \ref{fig:fig6}b).
We quantified the generalization performance to sample efficiency ratio as the generalization inefficiency, finding that after learning only 7 C-PRO contexts, vanilla ANNs generalized worse than pretrained ANNs (SFig. \ref{fig:fig6}c).
These findings support the notion that pretraining can simultaneously improve compositional generalization and sample efficiency.

\subsection{Convergent hierarchy of abstract representations in humans and ANNs}

\begin{figure}[h]
  \includegraphics[width=\linewidth]{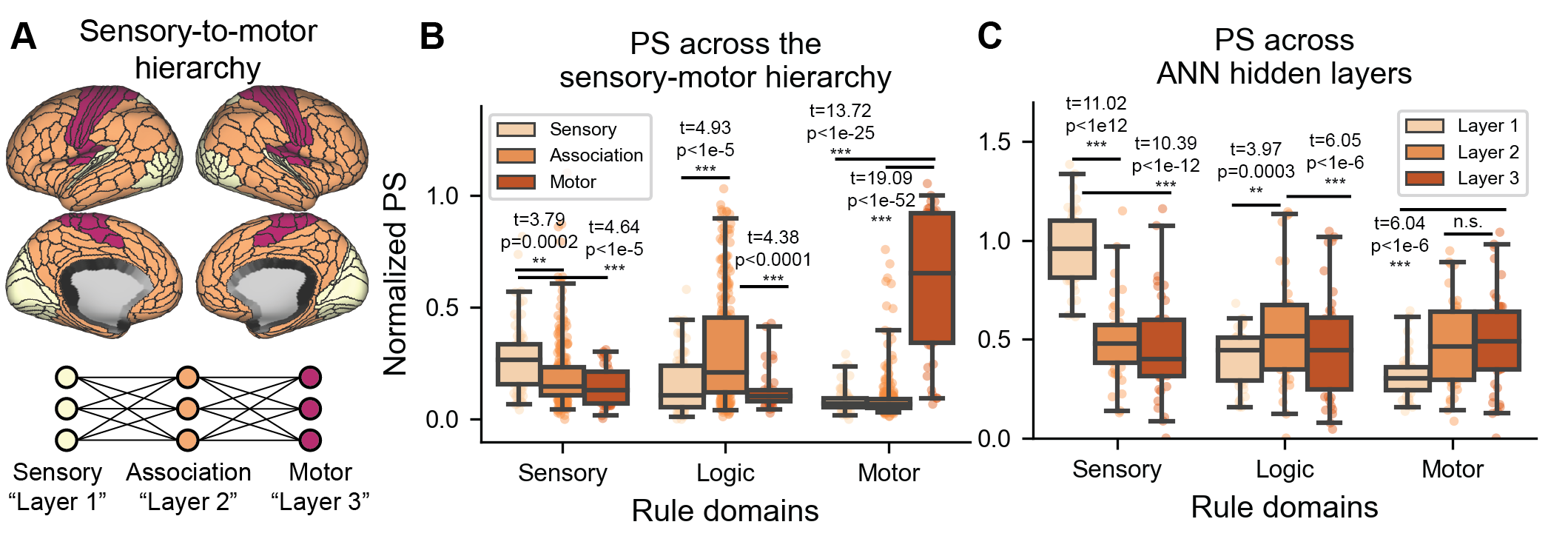}
  \caption{a) A discretized sensory-to-motor hierarchy (see SFig. \ref{fig:figs_networkpartition} for discretization details). b) We computed the normalized PS (i.e., the PS of each brain region normalized by the maximum PS across all regions) for each rule domain across the discretized cortical systems. c) Same analysis as in b), but using the PS found in each ANN hidden layer.}
   \label{fig:fig7}
\end{figure}

Analysis of human fMRI data revealed that content-specific abstraction was spatially heterogeneous across cortex. Recent neuroscience work has identified hierarchical gradients that organize along a sensory-to-motor output axis in both resting-state \citep{margulies_situating_2016} and multi-task fMRI data \citep{ito_multi-task_2021}. We therefore sought to quantify PS across this sensory-to-motor hierarchy in fMRI data, and compare it to PS changes in the feedforward hierarchy (i.e., layer-depth) in ANNs. We focused our analyses on the Combined pretrained model (which incorporates both Primitives and Simple task pretraining) due to its excellent zero-shot generalization (Fig. \ref{fig:fig5}b). In addition, we extended our model to include three hidden layers to make it easier to compare PS of different hidden-layer depths to the three cortical systems of interest: sensory, association, and motor systems (Fig. \ref{fig:fig7}a).

We measured the PS for each rule domain for sensory, association, and motor systems. Sensory rule PS was highest in the sensory system, logic rule PS was highest in association systems, and motor rule PS was highest in the motor system (Fig. \ref{fig:fig7}b). To observe whether similar hierarchical PS organization emerged in ANNs, we used the Combined pretrained model with three hidden layers, and plotted PS as a function of ANN depth. Since our ANN transformed sensory inputs into motor outputs, we analogized each ANN layer to the sensory, association, and motor cortical systems (Fig. \ref{fig:fig7}a). We found a similar pattern in the ANN: sensory PS peaked in the first hidden layer; logic PS peaked in the second hidden layer; and motor PS peaked in the last two hidden layers (Fig. \ref{fig:fig7}d). We corroborated these findings using a continuous sensory-motor hierarchical gradient map (without discretization) (SFig. \ref{fig:figs_continuousPS}-\ref{fig:figs_networkpartition}). (In addition, see SFig. \ref{fig:figs_pairwisePS_logic}-\ref{fig:figs_pairwisePS_motor} for PS scores for all possible rule dichotomies in human fMRI data and ANN activations.) These findings suggest that abstraction emerges as a function of rule-dependent specialization and hierarchical organization. 

\section{Discussion, Limitations, Conclusions}

We provide empirical support for the role of compositionality in human generalization, and implicate abstract representations as its neural implementation. 
In classic ANNs, which are known to perform poorly during systematic generalization \citep{hupkes_compositionality_2020,dekker_determinants_2022}, we found that computationally cheap pretraining paradigms embedded abstract representations that led to human-like generalization performance and sample efficient learning.
When mapping abstract representations across cortex and ANN layers, we found converging patterns of rule-specific abstractions from early sensory areas/layers to late motor areas/layers across human and ANN hierarchies. 
These results reveal the hierarchical organization of content-specific abstractions in the human brain and ANNs, while revealing the impact of these abstractions for compositional generalization in models.

Our pretraining approach directly leverages knowledge of task structure to design pretraining routines that embed task biases into ANNs. 
Despite the sample efficiency of this approach, this pretraining approach requires the initial overhead of designing paradigms useful for downstream learning. 
A related approach that similarly requires prior knowledge of task structure is ``representational backpropagation'' -- a regularization approach that aims to produce an idealized hidden representation \citep{kepple_curriculum_2021}. 
However, there are other inductive bias approaches that do not require prior task knowledge. 
One approach constrains ANNs to produce abstract task representations by initializing ANN weights from a low-norm distribution \citep{flesch_orthogonal_2022}. 
However, initializing ANN weights in this regime is computationally costly. Another approach is to initialize networks with built-in modular structures to facilitate the re-use of network modules across tasks \citep{marton_efficient_2021,rosenbaum_routing_2017}. 
However, exactly how such networks disentangle representations has not yet been explored. Nevertheless, all these approaches are complementary to each other. 
It will be important for future work to assess how these approaches may synergistically interact to optimize for sample-efficient generalization in multi-task settings.
	
Though we provide comprehensive evidence of the role of abstraction in compositional generalization, there are several limitations in the present study that future research can explore. 
We found that the spatial topography of abstract representations was highly content-dependent. 
However, analyses were limited to cross-context manipulations of limited rule types (sensory, logic, and motor gating), without addressing the organization of other task components (e.g., reward or stimuli). 
Thus, future studies can explore how brains and ANNs represent the abstraction of other task components. 
Second, though we were able to explore cross-context generalization across 64 contexts -- significantly more than previous empirical studies in neuroscience -- cross-context analysis was limited to a single task type (i.e., the C-PRO paradigm). 
It will be critical to see the organization of abstraction in multi-task settings that go beyond 64 contexts. 
Finally, our ANN modeling approach revealed the computational benefits of pretraining. 
It will be important for future work to benchmark sample efficiency and generalization performance against other training paradigms (e.g., in continual learning and/or meta-learning settings; \citep{hadsell_embracing_2020,wang_prefrontal_2018}). 

In conclusion, we characterized a convergent hierarchical organization of abstract representations across the human cortex and in ANNs using a 64-context paradigm, and provided insight into the impact of abstract representations on generalization performance.
Overall, we found that simple pretraining tasks efficiently embed abstract representations into ANNs, leading to improved systematic generalization similar to human behavior. 
These findings provide a human-centric benchmark from which to understand compositional generalization in ANNs, paving the way for greater interpretability of compositionality in ANNs. 
Importantly, investigating compositionality through a human-centric framework (e.g., by benchmarking ANNs against humans in the same task) creates a concrete target for interpreting the strengths and limitations of compositionality in ANNs. 
We hope these findings inspire further investigations into the comparison of compositionality in humans and ANNs.

\begin{ack}
T.I. acknowledges a research internship sponsored by IBM Research and a Swartz Foundation postdoctoral fellowship from Yale University. This project was supported in part by the US National Institutes of Health, under awards K99-R00 MH096901 and R01 MH109520 (M.W.C.). The content is solely the responsibility of the authors and does not necessarily represent the official views of any of the funding agencies.

\end{ack}

%
%
%
%
\small

\bibliographystyle{plainnat}
\bibliography{mybib,additional}

\section*{Checklist}


\begin{enumerate}

\item For all authors...
\begin{enumerate}
  \item Do the main claims made in the abstract and introduction accurately reflect the paper's contributions and scope?
    \answerYes{}
  \item Did you describe the limitations of your work?
    \answerYes{see Section 4.}
  \item Did you discuss any potential negative societal impacts of your work?
    \answerNA{}
  \item Have you read the ethics review guidelines and ensured that your paper conforms to them?
    \answerYes{All human data has been de-identified and been publicly made available.}
\end{enumerate}

\item If you are including theoretical results...
\begin{enumerate}
  \item Did you state the full set of assumptions of all theoretical results?
    \answerNA{}
        \item Did you include complete proofs of all theoretical results?
    \answerNA{}
\end{enumerate}

\item If you ran experiments...
\begin{enumerate}
  \item Did you include the code, data, and instructions needed to reproduce the main experimental results (either in the supplemental material or as a URL)?
    \answerYes{Public data URL is written in the text (Section 2.1)}
  \item Did you specify all the training details (e.g., data splits, hyperparameters, how they were chosen)?
    \answerYes{}
        \item Did you report error bars (e.g., with respect to the random seed after running experiments multiple times)?
    \answerYes{Applicable to Figures 1, 5, 6, 7}
        \item Did you include the total amount of compute and the type of resources used (e.g., type of GPUs, internal cluster, or cloud provider)?
    \answerYes{}
\end{enumerate}

\item If you are using existing assets (e.g., code, data, models) or curating/releasing new assets...
\begin{enumerate}
  \item If your work uses existing assets, did you cite the creators?
    \answerYes{Used previously published data \citep{ito_cognitive_2017}}
  \item Did you mention the license of the assets?
    \answerYes{Published under a CC0 license, mentioned in section A.1.}
  \item Did you include any new assets either in the supplemental material or as a URL?
    \answerNo{}
  \item Did you discuss whether and how consent was obtained from people whose data you're using/curating?
    \answerYes{Yes, see section A.1.}
  \item Did you discuss whether the data you are using/curating contains personally identifiable information or offensive content?
    \answerYes{Yes, see section A.1. All data was previously de-identified.}
\end{enumerate}

\item If you used crowdsourcing or conducted research with human subjects...
\begin{enumerate}
  \item Did you include the full text of instructions given to participants and screenshots, if applicable?
    \answerYes{Yes -- see Figure 1A}
  \item Did you describe any potential participant risks, with links to Institutional Review Board (IRB) approvals, if applicable?
    \answerYes{IRB approval was mentioned in section A.1}
  \item Did you include the estimated hourly wage paid to participants and the total amount spent on participant compensation?
    \answerNo{This was a previously published dataset.}
\end{enumerate}

\end{enumerate}


\newpage

\appendix

\section{Supplementary Methods}

\subsection{Participants}

The following description is quoted with citation from a previous study that used the same dataset \citep{ito_constructing_2022}. The dataset was publicly published under a CC0 license, and is publicly available (\url{https://openneuro.org/datasets/ds003701}). All participant data on the public repository has been de-identified.

Data were collected from 106 human participants across two different sessions (a behavioral and an imaging session). Technical error during MRI acquisition resulted in removing six participants from the study. Four additional participants were removed from the study because they did not complete the behavior-only session. fMRI analysis was performed on the remaining 96 participants (54 females). All participants gave informed consent according to the protocol approved by the Rutgers University Institutional Review Board. The average age of the participants that were included for analysis was 22.06, with a standard deviation of 3.84.


\subsection{C-PRO task paradigm -- additional details}

The C-PRO cognitive paradigm permutes specific task rules from three different rule domains (logical decision, sensory semantic, and motor response) to generate dozens of novel and unique task contexts. Visual stimuli included either horizontally or vertically oriented bars with either blue or red coloring. Simultaneously presented auditory stimuli included continuous (constant) or non-continuous (non-constant, i.e., ``beeping'') tones presented at high (3000Hz) or low (300Hz) frequencies. A given task context could be presented with 256 unique stimulus combinations. This is because a given task context was presented with two sequentially presented audiovisual stimuli, where each audiovisual stimulus varied in four dimensions: color (red/blue), orientation (vertical/horizontal), pitch (high/low), continuity (continuous/beeping). This led to $2^8=256$ possible stimulus combinations. The paradigm was presented using E-Prime software version 2.0.10.353 \cite{schneider_e-prime:_2002}.

Each rule domain (logic, sensory, and motor) consisted of four specific rules, while each task context was a combination of one rule from each rule domain. A total of 64 unique task contexts (4 logic rules x 4 sensory rules x 4 motor rules) were possible, and each unique task set was presented twice for a total of 128 task miniblocks. This meant that there were $256*64=16384$ unique trials (i.e., context-stimulus) combinations. Identical task contexts were not presented in consecutive blocks. Each task miniblock included three trials, each consisting of two sequentially presented instances of simultaneous audiovisual stimuli. A task block began with a 3925 ms encoding screen (5 TRs), followed by a jittered delay ranging from 1570 ms to 6280 ms (2-8 TRs; randomly selected). Following the jittered delay, three trials were presented for 2355 ms (3 TRs), each with an inter-trial interval of 1570 ms (2 TRs). A second jittered delay followed the third trial, lasting 7850 ms to 12560 ms (10-16 TRs; randomly selected). A task block lasted a total of 28260 ms (36 TRs). Subjects were trained on four of the 64 task contexts for 30 minutes prior to the fMRI session. The four practiced rule sets were selected such that all 12 rules were equally practiced. There were 16 such groups of four task sets possible, and the task sets chosen to be practiced were counterbalanced across subjects. Subjects' mean performance across all trials performed in the scanner was 84\% (median=86\%) with a standard deviation of 9\% (min=51\%; max=96\%). All subjects performed statistically above chance (25\%).

\subsection{Analysis of human task performance data}

The corresponding results described in this section can be found in Fig. \ref{fig:fig1}.

We calculated the average accuracy for each task miniblock (comprising three task trials). Note that each task miniblock had the same task context (three-rule combination) for all three trials. This resulted in 128 task accuracy scores for every subject. Task rule contexts were then sorted into three categories separately for every subject: practiced, 2-rule overlap, and 1-rule overlap. Practiced task contexts were defined as the four task contexts that were used to train participants on the C-PRO task outside of the MRI scanner. 2-rule overlap tasks were task contexts that had at least 2 of the same rules overlapping with the previously seen practiced tasks, and 1-rule overlap tasks were tasks with only a 1-rule overlap with practiced tasks. Note that there were no 0-rule overlap tasks, since subjects were trained on every rule prior to the test session. Moreover, there were no rule overlaps across practiced task contexts. Finally, every participant was provided with a randomly selected set of practiced task contexts. Behavioral accuracy was computed for every task context group for every subject (Fig. \ref{fig:fig1}b).

We next evaluated the successive miniblock performance of each rule presented in a novel task context during the fMRI scanning session (see Fig. \ref{fig:fig1}c-d). This would capture a participant's ability to use a previously seen rule in a novel context as a function of the number of times it used the rule previously. The performance of an individual rule (e.g., ``BOTH'') was calculated separately per participant as a function of each novel context seen. Thus, performance of each rule was calculated for exactly 15 miniblocks (each rule was presented 16 times, including the practiced miniblock). For each rule, we then fit a linear regression model to assess whether performance (dependent variable) could be calculated as function of miniblock presentation (independent variable) with a positive coefficient (i.e., increasing slope). A significant positive increase of performance (increasing positive performance) was tested for significance by assessing the p-value of the beta coefficient (p$<$0.05 threshold). This captured compositional learning -- re-using a task rule (despite use in a novel context) indicated that participants were learning to use previously learned rules in out-of-set novel contexts.

\subsection{fMRI acquisition}

The following fMRI acquisition details are taken from a previous study that used the same data set \citep{ito_constructing_2022}.

Whole-brain multiband echo-planar imaging (EPI) acquisitions were collected with a 32-channel head coil on a 3T Siemens Trio MRI scanner with TR=785 ms, TE=34.8 ms, flip angle=55$^{\circ}$, Bandwidth 1924/Hz/Px, in-plane FoV read=208 mm, 72 slices, 2.0 mm isotropic voxels, with a multiband acceleration factor of 8. Whole-brain high-resolution T1-weighted and T2-weighted anatomical scans were also collected with 0.8 mm isotropic voxels. Spin echo field maps were collected in both the anterior to posterior direction and the posterior to anterior direction in accordance with the Human Connectome Project preprocessing pipeline \cite{glasser_human_2016}. A resting-state scan was collected for 14 minutes (1070 TRs), prior to the task scans. Eight task scans were subsequently collected, each spanning 7 minutes and 36 seconds (581 TRs). Each of the eight task runs (in addition to all other MRI data) were collected consecutively with short breaks in between (subjects did not leave the scanner).

\subsection{fMRI preprocessing}

The following details are quoted with citation from a previous study that used the same preprocessing scheme \citep{ito_constructing_2022}.

Resting-state and task-state fMRI data were minimally preprocessed using the publicly available Human Connectome Project minimal preprocessing pipeline version 3.5.0. This pipeline included anatomical reconstruction and segmentation, EPI reconstruction, segmentation, spatial normalization to standard template, intensity normalization, and motion correction. After minimal preprocessing, additional custom preprocessing was conducted on CIFTI 64k grayordinate standard space for vertex-wise analyses using a surface based atlas \citep{glasser_human_2016}. This included removal of the first five frames of each run, de-meaning and de-trending the time series, and performing nuisance regression on the minimally preprocessed data \citep{ciric_benchmarking_2017}. We removed motion parameters and physiological noise during nuisance regression. This included six motion parameters, their derivatives, and the quadratics of those parameters (24 motion regressors in total). We applied aCompCor on the physiological time series extracted from the white matter and ventricle voxels (5 components each extracted volumetrically) \citep{behzadi_component_2007}. We additionally included the derivatives of each component time series, and the quadratics of the original and derivative time series (40 physiological noise regressors in total). This combination of motion and physiological noise regressors totaled 64 nuisance parameters, and is a variant of previously benchmarked nuisance regression models \citep{ciric_benchmarking_2017}.

\subsection{fMRI activation estimation}

We performed a within-subject task GLM on the vertex-wise fMRI time series to estimate task rule-related activations on the CIFTI grayordinate space. To extract task activations for each task block, we performed a beta series regression on every task miniblock \citep{rissman_measuring_2004}. Specifically, we fit an independent regressor to every encoding period (3925ms, 5 TRs), resulting in 128 task regressors in total. Fitting regressors on the encoding period was done primarily to isolate rule representations rather than the actual trial (stimulus-response) period. Each regressor was a boxcar function that was a vector of 0s, except for the specified encoding period. This boxcar function was then convolved with the SPM canonical hemodynamic response function \citep{friston_statistical_1994}. A single activation estimate (beta coefficient) was extracted for every encoding block at every surface vertex.

\subsection{ANN construction and batch training}

The primary ANN architecture was comprised of two hidden layers, each with 128 units. The output layer was comprised of four units that corresponded to each motor response. The ANN transformed the trial input vector into a 4-element response vector with the equation $ Y = f_{ReLU}(X_{h}W_{h}+b_{h}) $, where $Y$ corresponds to the output vector, $W_{h}$ is the weight matrix from the last hidden layer to the output layer, $X_{h}$ is the activation vector of the last hidden layer, and $b_{h}$ is the bias vector. The hidden unit activation vectors were defined as $ X_{i} = f_{ReLU}(X_{i-1}W_{i-1} + b_{i-1})$, where $X_{i}$ is the activation vector for layer $i$. Weights and biases were initialized from a uniform distribution $U(-\sqrt{1/k},\sqrt{1/k})$, where $k$ is the number of input features from the previous layer.

The ANN was optimized by minimizing the cross entropy between the outputs and the correct target output (a one-hot vector). Optimization was performed using Adam with a learning rate of 0.001 \citep{kingma_adam_2017}. During training, we used dropout (with probability 0.2 from input to hidden, and 0.5 within hidden layers). 

Training on the C-PRO task was performed in a sequential learning paradigm. Initially, an arbitrary set of four practiced contexts were selected with the constraint that no set of four practiced contexts had any overlapping rules. (This was randomly selected across different ANN initializations). Then, novel task contexts were incrementally added (by 1) into the set of training contexts, and performance (and ANN analysis) was performed after the addition of each task context. Training stopped once all 64 task contexts had been fully trained on. We used batch training. Each batch contained a single task context with all possible stimulus (256) combinations. Thus, each batch contained 256 trials in total. 

To stop training we set criteria for two different experiments. The first experiment required that performance on each task context (each batch) in the training set achieved a baseline performance accuracy or better (90\%). Once this criterion was satisfied, a novel task context would be added into the training. The second experiment kept fix the number of batches/gradient steps each task context took prior to adding a new task context. This was set to 200 gradient steps per task context.

\subsection{ANN pretraining description}

There were two pretraining paradigms: Primitives pretraining and Simple task pretraining. 	

Primitives pretraining focused on training individual rules within each rule domain, enabling ANNs to learn individual rule Primitives (e.g., the notion of what ``RED'' is). This is consistent with how humans enter the full C-PRO experiment -- most participants already know what the primitives ``RED'', ``LEFT MIDDLE'', or ``BOTH'' refer to. While C-PRO task contexts activated 3 rules out of 12 possible task rules, Primitives tasks only activated 1/12 rules. Primitives pretraining was performed for each of the rule domains separately. 

Motor Primitives pretraining involved making motor responses given a motor rule. This involved activating one motor rule in the input vector at a time. If the ``LEFT MIDDLE'' rule was activated, a LEFT MIDDLE output response would be expected. In addition, if the ``NOT'' unit was activated in conjunction with the ``LEFT MIDDLE'' rule, then the ``LEFT INDEX'' output response would be expected, which is the analogous rule instruction that participants received prior to performing the C-PRO task.

Sensory Primitives pretraining involved making True/False statements on whether a specific sensory stimulus feature was presented. This involved activating one sensory rule and one sensory stimulus feature at a time. For example, for the sensory rule ``RED'', either a RED/BLUE sensory stimulus would activate. During pretraining, we included two additional output units that corresponded to True/False units. If a ``RED'' stimulus was presented in conjunction with ``RED'' sensory rule, the output should be ``True''; otherwise, ``False''. In addition, we also presented sensory rules with the ``NOT'' negation. In other words, if the rule was ``NOT RED'' and the sensory stimulus presented was RED, then the network should produce ``False''.

Logic Primitives pretraining involved learning the abstract logical relations between different logic rules. The logic rules ``BOTH'' and ``EITHER'' could be equivalently interpreted as ``AND'' and ``OR'' logic rules, respectively. The ``NOT BOTH'' and ``NEITHER'' rules were analogous to the negations of those rules. This was operationalized by presenting two stimuli from a specific feature domain (e.g., color). Using stimuli from the color feature domain as an example, for ``BOTH'' Primitive training, if RED-then-RED stimuli were presented, this would result in a ``True'' boolean. In contrast, if RED-then-BLUE stimuli were presented, then this would result in a ``False'' boolean. This intuition was derived from conditional logic (where ``BOTH'' is equivalent to the concept of ``SAME''), where the statement $x == x$ is True. In contrast, ``EITHER'' (i.e., ``OR'') was coded as ``True'' for any combination features of a color stimuli. The negations ``BOTH'' and ``EITHER'' resulted in the negation of the produced boolean (i.e., ``NEITHER'' produced ``False'' for all stimulus pairs). The most critical point, however, is that the responses for all logic rules were distinct among each other, and described equivalent logical relations between rules. Previous theoretical work in cognitive science suggests that symbolic computation emerges by learning the relational representations between symbolic operations \citep{piantadosi_computational_2020}. This Logic Primitives pretraining approach captures the relations amongst symbolic operations.

Simple task pretraining involved combining two-rule task context, rather than the full three-rule task context in the C-PRO task. We designed two variants of simple tasks: a sensorimotor task that combined a sensory and motor rule, and a logical-sensory task that combined a logic and sensory rule. The sensorimotor task activated a sensory rule, motor rule, and one sensory stimulus unit. The response output was a motor response. For example, if the rules were ``RED'' and ``LEFT MIDDLE'', and the stimulus was ``RED'', ANNs were taught to respond with the Left Middle response unit. If the stimulus was anything other than ``RED'', then ANNs were taught to respond with the Left Index response unit. The sensorimotor task was used to teach ANNs simple mappings between sensory input and motor responses in the simplest possible paradigm.

The logical-sensory task was designed to teach ANNs logical inferences over sensory stimuli in the simplest possible manner. Like the sensorimotor task, the logical-sensory task included two rules: a logic and sensory rule. However, unlike the sensorimotor task, it activated two stimulus units from the same feature domain (e.g., the color domain). For example, for the logic rule was ``BOTH'' and sensory rule ``RED'', if the first stimulus was red and second stimulus red, the ANN would be taught to respond with the True output unit. If any of the stimuli were not red (e.g., blue), then the ANN would be taught to respond with the False output unit. 

Primitives pretraining was always performed prior to Simple task pretraining, except for SFig. \ref{fig:figs_ann_reversepretrained}, which investigated the effect of reversing the order of pretraining tasks. 
Pretraining procedures were blocked together, such that all conditions within the Primitives pretraining paradigm (i.e., Logic, Sensory, and Motor primitives pretraining) were trained until all three tasks achieved 99\% accuracy. Simple task pretraining was subsequently performed until both Logical-Sensory and Sensory-Motor tasks were performed at 99\% accuracy. This ordering is consistent with prior work, suggesting that ANNs are more sample efficient when transitioning from easier to more difficult tasks \citep{saglietti_analytical_2021}. 
Conditions within each pretraining protocol were interleaved \citep{fusi_neural_2007}, ensuring that catastrophic forgetting was not an issue, as is common in continual learning paradigms \citep{flesch_comparing_2018}.
We also performed a simple control experiment demonstrating that when pretraining was reversed in the Combined condition (i.e., Simple task pretraining followed by Primitives pretraining), generalization performance was reduced to chance (SFig. \ref{fig:figs_ann_reversepretrained}). This suggests that the ordering of pretraining paradigms is crucial for generalization performance and sample efficiency, which future work should explore.

After pretraining, the additional True/False output units were lesioned from the network. 

\subsection{ANN analysis}

Analysis of ANNs was carried out in a similar manner to how empirical fMRI data and behavior was analyzed. ANN analysis was performed to infer how the structure of their internal representations was associated with task generalization performance and sample efficiency. Task generalization performance was calculated as the performance on novel contexts (i.e., novel recombinations of task rules). This was independent of whether or not ANNs had learned/seen individual rules previously. To estimate task sample efficiency, we calculated the number of trial samples required to achieve a baseline task accuracy percentage (on contexts in the training set). Note that for a fixed number of trial samples, the number of gradient steps were the same (batch sizes were always the same). Generalization inefficiency was measured as the ratio of the number of samples trained on (normalized between 0 and 1) and the generalization performance on novel contexts (normalized to 0 to 1 + a fixed constant).
	
PS in ANNs was calculated separately from the training procedure. 
Since we were only interested in the PS of rule representations, only input units associated with the task rules were activated, while stimulus inputs were set to 0.
This ensured that the hidden activations were not contaminated by stimulus-related activations when calculating PS in the hidden layers.
Otherwise, PS in ANNs was calculated in a similar manner to the empirical fMRI data, where the spatial features (dimensions) were the units within a given hidden layer (like voxels within a brain parcel).

In addition, Supplementary Figures \ref{fig:figs_pairwisePS_logic}, \ref{fig:figs_pairwisePS_sensory}, and \ref{fig:figs_pairwisePS_motor} illustrate that pairwise PS scores for all pairs of rule dichotomies in both ANNs and human fMRI data.

\subsection{Computing resources}

fMRI analyses were carried out on a local server with 24 cores and 320GB of RAM. ANN training, while not required, was performed on an NVIDIA P100 GPU. A single ANN initialization can be successfully trained on a CPU in under 2 minutes.

\section{Supplementary Figures}

\setcounter{figure}{0}  
\renewcommand{\figurename}{Supplementary Figure}

\begin{figure}[h]
  \includegraphics[width=\linewidth]{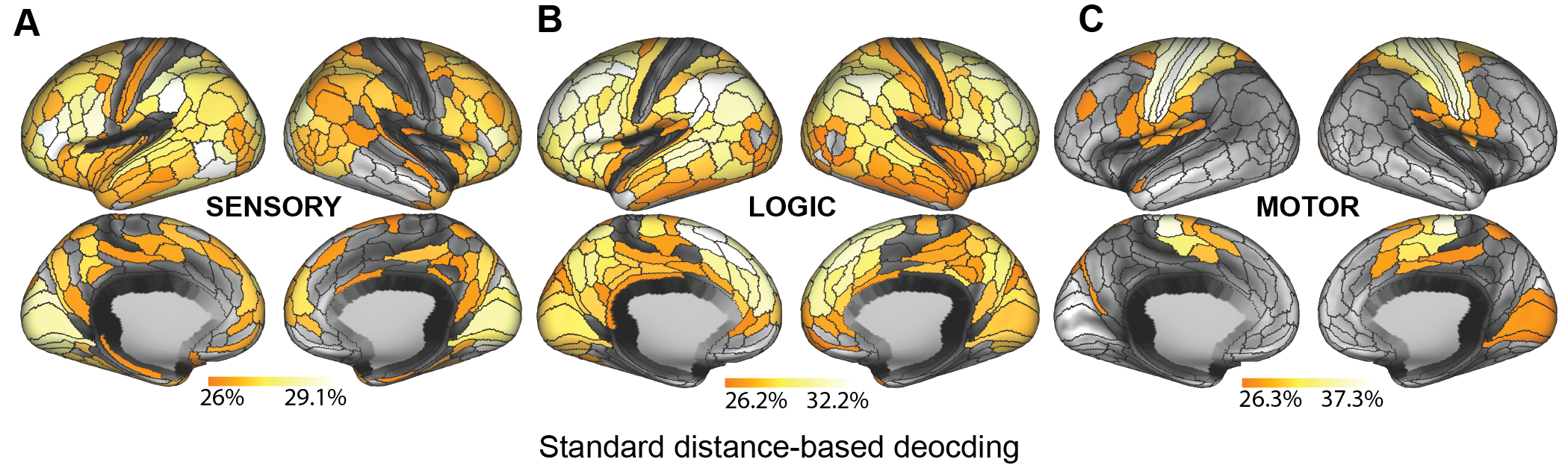}
  \caption{Standard decoding of multivariate activation patterns reveals more distributed patterns of decodability than identified with PS. These findings suggest that of the regions that contain task information, only a subset of these regions contain abstract representations. a) Sensory, b) logic, and c) motor rule decoding at the group level (n=96). Decoding was performed using a distance-based classifier (Pearson correlation), and significance was assessed using a binomial test against chance (25\%). Significance was assessed using multiple comparisons-corrected threshold (False Discovery Rate) of p$<$0.05.}
   \label{fig:figs_standarddecoding}
\end{figure}



\begin{figure}[h]
  \includegraphics[width=\linewidth]{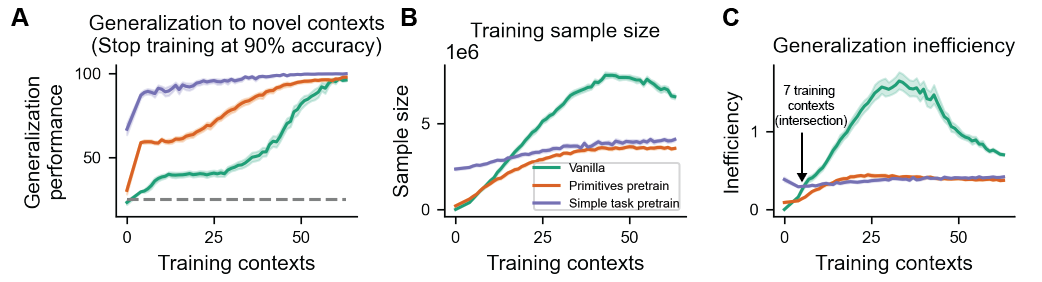}
  \caption{a) Generalization performance on novel contexts after training on n/64 contexts (x-axis). b) Number of total trials/samples shown to each model type. c) The generalization inefficiency of each model. Generalization inefficiency was measured as the ratio of the number of samples shown (normalized between 0 and 1) and the performance on novel tasks (normalized to 0 to 1 + a fixed constant). See SFig. \ref{fig:figs_ann_fixedepoch} for model performance using a fixed number of samples per trained context.} 
  \label{fig:fig6}
\end{figure}

\begin{figure}[h]
  \includegraphics[width=\linewidth]{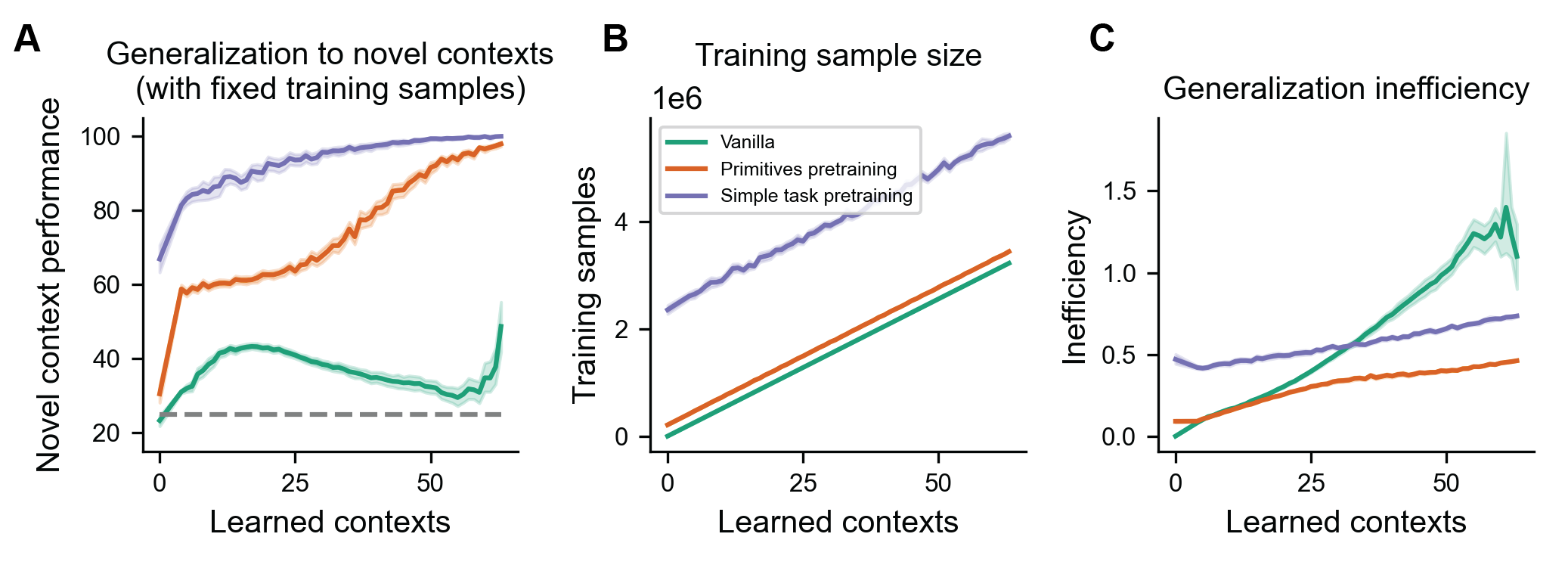}
  \caption{Novel context performance as function of learned contexts in the C-PRO task in the ANN model. Similar to SFig. \ref{fig:fig6}, but we trained the model using a fixed number of samples per learned context (200 epochs per context) and evaluated generalization performance. a) We systematically trained each type of ANN on a range from 0-63 of the C-PRO task contexts, and assessed its generalization on the remaining (novel) contexts. We trained the model on a fixed number of samples per learned context. Learned contexts were sequentially introduced, and novel context performance was assessed on the remaining (excluded from training) contexts. b) The number of total trials/samples shown to each model type. Since the number of samples shown to the model for each context was fixed, training samples linearly increased as a function of learned contexts. c) As in SFig. \ref{fig:fig6}c, we measured the generalization inefficiency of each model. Thus, while the Vanilla model was initially more efficient, pretrained models quickly learned more efficiently as evidenced by better generalization accuracies with fewer samples. }
   \label{fig:figs_ann_fixedepoch}
\end{figure}

\begin{figure}[h]
  \centering\includegraphics[width=4.54in]{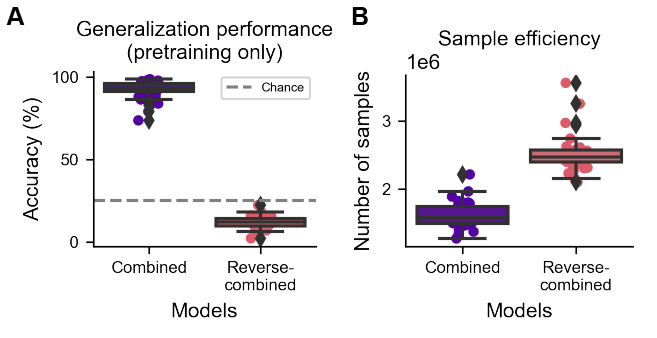}
  \caption{Order of pretraining paradigms drastically affects generalization performance and sample efficiency. In the Combined pretraining regimen, we first implemented Primitives pretraining followed by Simple task pretraining. However, prior curriculum learning research suggests that the order of task learning can significantly impact generalization performance \citep{saglietti_analytical_2021}. Thus, we compared the Combined pretraining protocol to a Reverse-combined protocol, where Simple task pretraining was performed prior to Primitives pretraining. a) We found that generalization performance on the unseen C-PRO tasks was significantly impaired in the Reverse-combined condition. b) We also measured the sample efficiency of only the pretraining trials. We found that Reverse-combined pretraining required more samples than Combined pretraining, despite the stopping criterion for each pretraining task remaining the same. (ANNs were required to perform either Primitives pretraining or Simple task pretraining with at least 99\% accuracy prior to moving on to the next pretraining task.) This suggests that ordering of pretraining tasks can significantly impact the learning and generalization dynamics of simple ANNs.}
   \label{fig:figs_ann_reversepretrained}
\end{figure}

\begin{figure}[h]
  \includegraphics[width=\linewidth]{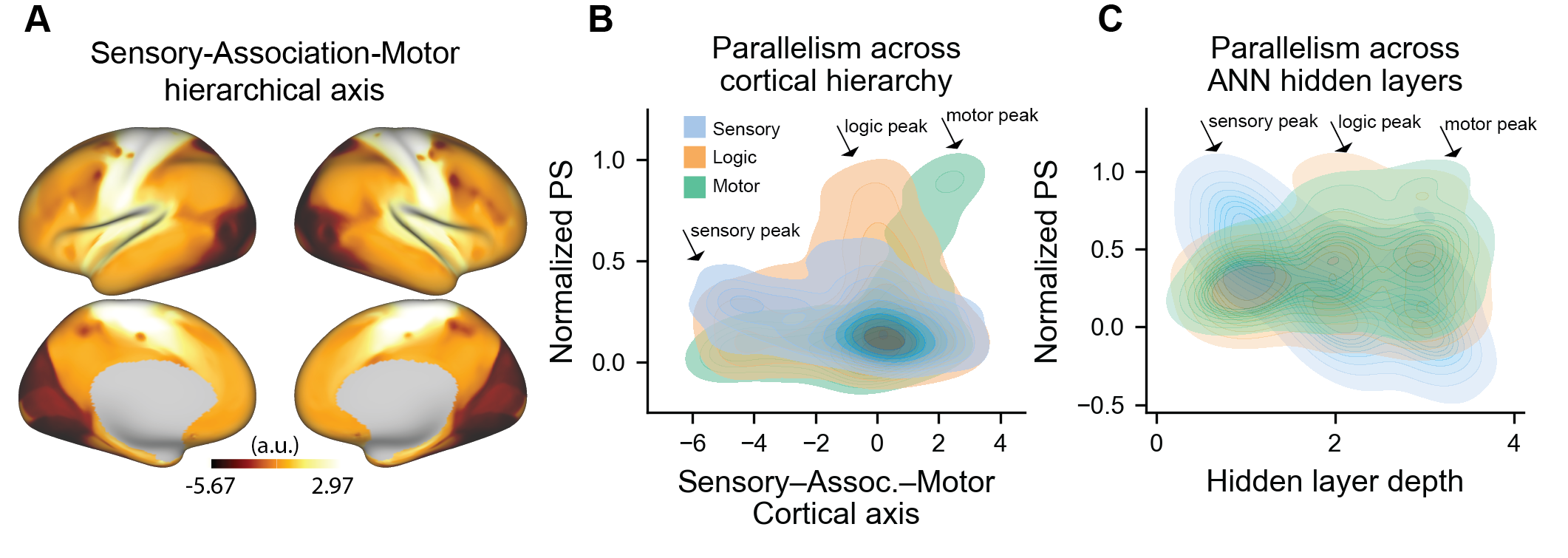}
  \caption{Convergent hierarchical organization of abstract representations in humans and ANNs. a) We compared the topographic differences in PS across cortex with a well-known sensory-to-motor hierarchical gradient identified during resting-state fMRI \citep{margulies_situating_2016}. b) We compared how content-specific abstractions (PS) differed across this hierarchical gradient, finding that sensory rule abstractions were highest in the lower part of the gradient (sensory systems), logic rule abstractions were highest across association cortex, and motor rule abstractions were highest across motor cortex. c) For an analogous analysis, we plotted how domain-specific abstraction differed across different hidden layer depths in the pretrained ANN with 3 hidden layers. We identified similar patterns of parallelism across the 3 hidden layers for each of the rule domains.}
   \label{fig:figs_continuousPS}
\end{figure}

\begin{figure}[h]
  \includegraphics[width=\linewidth]{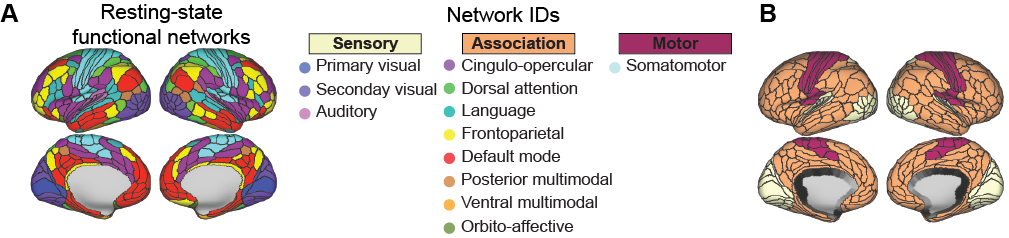}
  \caption{a) A network partition of cortical parcels using resting-state fMRI \citep{ji_mapping_2019}. b) To match the number of layers in the ANN, we created three discretized systems based on the functional network partition -- sensory, association, and motor -- that followed the sensory-to-motor hierarchy (Fig. \ref{fig:fig7}) \citep{margulies_situating_2016,ito_multi-task_2021}}
   \label{fig:figs_networkpartition}
\end{figure}

\begin{figure}[h]
  \includegraphics[width=\linewidth]{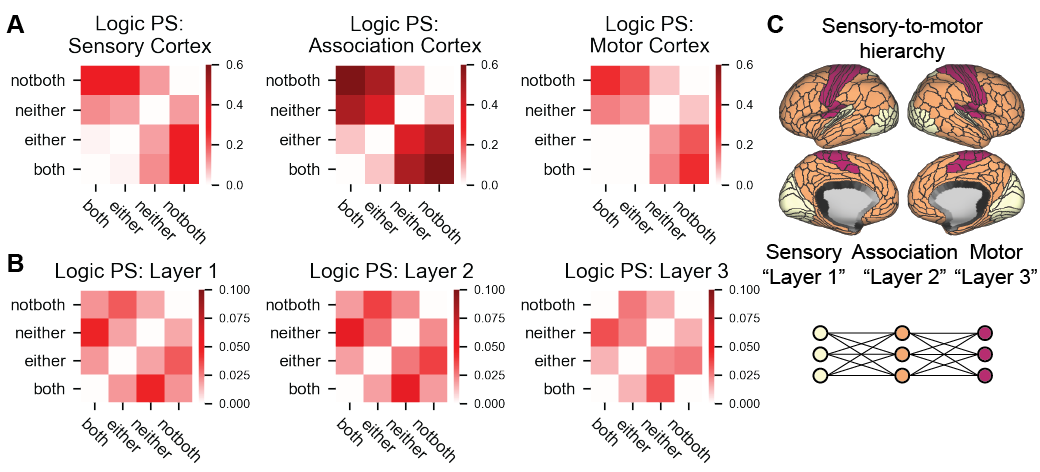}
  \caption{A pairwise comparison of PS for all possible dichotomies in the Logic rule condition. a) We computed the PS for every pairwise dichotomy in the Logic rule domain, and computed the average PS across brain regions within each cortical system (Sensory, Association, and Motor cortex). The average PS for each cortical system was computed by averaging the PS across all regions within that cortical system for every dichotomy. b) To compare how the pairwise dichotomies matched in pretrained ANNs (Combined pretrained only; no training on full C-PRO trials), we computed the PS for all dichotomies in each layer. In this particular experiment we used an ANN with three hidden layers to compare with Sensory, Association, and Motor cortical systems. c) The sensory, association, and motor cortical systems can be analogized to hidden layer depths in the ANN.}
   \label{fig:figs_pairwisePS_logic}
\end{figure}

\begin{figure}[h]
  \includegraphics[width=\linewidth]{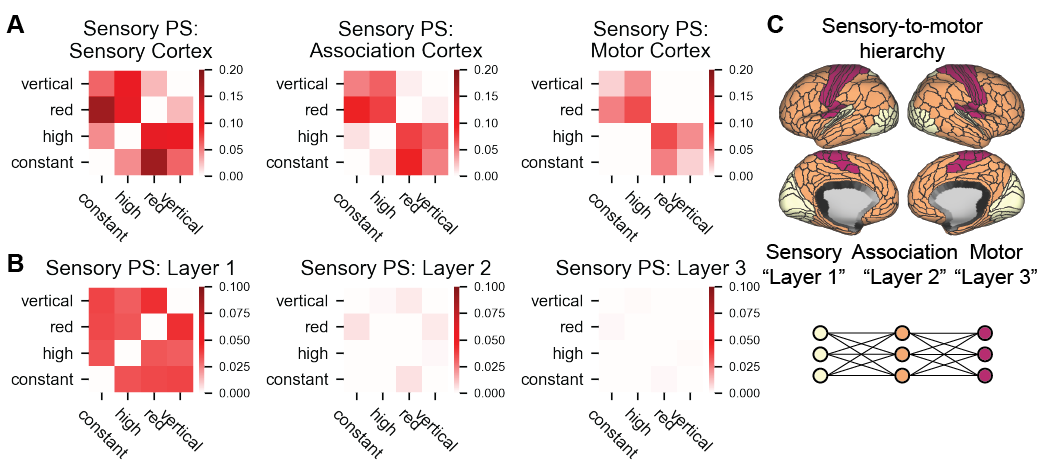}
  \caption{A pairwise comparison of PS for all possible dichotomies in the Sensory rule condition. a) We computed the PS for every pairwise dichotomy in the Sensory rule domain, and computed the average PS across brain regions within each cortical system (Sensory, Association, and Motor cortex). The average PS for each cortical system was computed by averaging the PS across all regions within that cortical system for every dichotomy. b) To compare how the pairwise dichotomies matched in pretrained ANNs (Combined pretrained only; no training on full C-PRO trials), we computed the PS for all dichotomies in each layer. In this particular experiment we used an ANN with three hidden layers to compare with Sensory, Association, and Motor cortical systems. c) The sensory, association, and motor cortical systems can be analogized to hidden layer depths in the ANN.}
   \label{fig:figs_pairwisePS_sensory}
\end{figure}

\begin{figure}[h]
  \includegraphics[width=\linewidth]{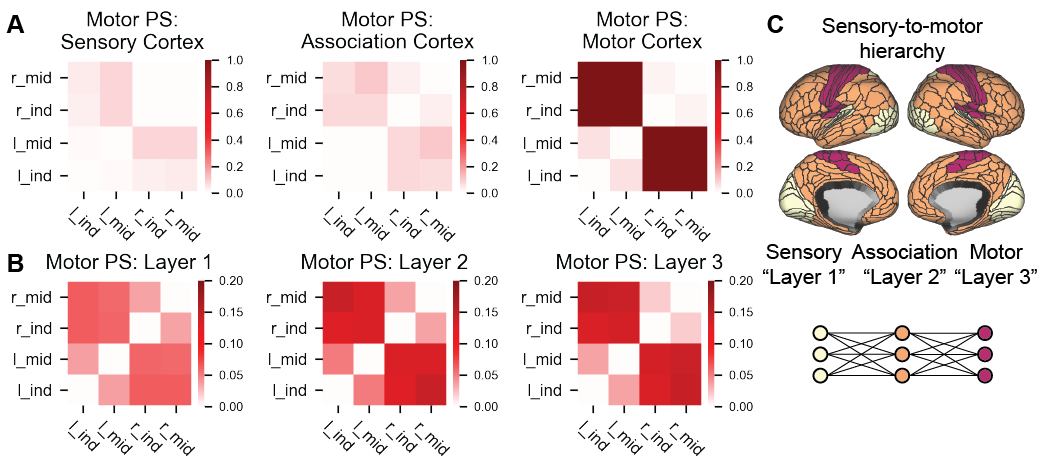}
  \caption{A pairwise comparison of PS for all possible dichotomies in the Motor rule condition. a) We computed the PS for every pairwise dichotomy in the Motor rule domain, and computed the average PS across brain regions within each cortical system (Sensory, Association, and Motor cortex). The average PS for each cortical system was computed by averaging the PS across all regions within that cortical system for every dichotomy. b) To compare how the pairwise dichotomies matched in pretrained ANNs (Combined pretrained only; no training on full C-PRO trials), we computed the PS for all dichotomies in each layer. In this particular experiment we used an ANN with three hidden layers to compare with Sensory, Association, and Motor cortical systems. c) The sensory, association, and motor cortical systems can be analogized to hidden layer depths in the ANN.}
   \label{fig:figs_pairwisePS_motor}
\end{figure}

\end{document}